\documentclass[12pt]{iopart}
\usepackage{iopams} 
\usepackage[usenames,dvipsnames]{xcolor} 
\begin{document}

\title[Over-spinning Kerr-Sen black holes with test fields]{Over-spinning Kerr-Sen black holes with test fields}

\author{Koray D\"{u}zta\c{s}}

\address{Physics Department, Eastern Mediterranean  University, Famagusta, North Cyprus, via Mersin 10, Turkey}
\ead{koray.duztas@emu.edu.tr}
\vspace{10pt}

\begin{abstract}
In this work we investigate validity of the weak form of the cosmic censorship conjecture in the interaction of Kerr-Sen black holes with neutral test fields. Previous studies of the Klein-Gordon equation on Kerr-Sen background imply that superradiance occurs for scalar fields. We show that scalar fields can overspin a nearly extremal black hole into a naked singularity, but the modes that could overspin an extremal black hole are not absorbed due to superradiance. From Kerr analogy one can naively expect superradiance to be absent for fermionic fields. In such a case overspinning becomes generic and also applies to extremal Kerr-Sen black holes. This robust violation of cosmic censorship cannot be fixed by backreaction effects which are ignored in this work. These results are analogous to the Kerr case.
\end{abstract}

%
\vspace{2pc}

\section{Introduction}
The singularity theorems developed by Penrose and Hawking in late sixties imply that the formation of singularities is inevitable as a result of gravitational collapse~\cite{singtheo}. This could lead to the breakdown of predictability and the deterministic nature of general relativity. Penrose proposed the cosmic censorship conjecture to circumvent this problem~\cite{ccc}. In its weak form (wCCC), the conjecture asserts that the gravitational collapse of a body always ends up in a black hole, rather than a naked singularity. The singularity is covered by an event horizon which disables its contact with distant observers. Thus, the observers in the asymptotically flat region do not encounter any effects propagating out of the singularity. In this way, predictability and the deterministic nature of general relativity is reassured at least in the asymptotically flat region.

If the gravitational collapse occurs in the way prescribed by Penrose and Hawking, trapped surfaces form as the body collapses beyond  its Schwarzschild radius. The gravitational collapse ends up as a black hole surrounded by an event horizon. Whether this can be generalised to include every form of gravitational collapse is still an open problem. No concrete proof could be given in decades. 

An alternative test of wCCC was suggested by Wald. In this approach one starts with an extremal or nearly extremal black hole and perturbs it with test particles or fields to check if it is possible to destroy the event horizon. In the first of these thought experiments Wald showed that particles with sufficient charge or angular momentum to overspin or overcharge an extremal Kerr-Newman black hole are not absorbed by the black hole~\cite{wald74}. Later, Hubeny came up with the idea of starting with a nearly extremal black hole instead of an extremal one. She showed that it is possible to overcharge a nearly extremal Reissner-Nordst\"{o}m black hole into a naked singularity~\cite{hubeny}. After Hubeny's attempt de Felice and Yu argued a naked singularity can evolve by the absorption of a neutral test particle by an extremal Reissner-Nordst\"{o}m black hole~\cite{f1} Similar tests of wCCC were also applied to  the black holes in Einstein-Maxwell theory ~\cite{saa,gao,magne,dilat}. The same approach was adapted to over-spin Kerr black holes by test particles~\cite{jacob}. Back-reaction effects were considered for some of these problems to prevent the horizon from being destroyed~\cite{back1,hu1,back2}. The validity of wCCC was also investigated for the asymptotically anti de-Sitter case \cite{vitor,zhang,rocha,gwak1,gwak2,btz} and higher dimensional black holes~\cite{higher,v1}. Recently Sorce and Wald has published new versions of the original thought experiment~\cite{wald1,wald2}

It is also possible to test wCCC in the case of test fields scattering off black holes.  Many thought experiments involving the perturbations of space-times by test fields were constructed in this vein~\cite{semiz,toth1,emccc,overspin,superrad,sh,duztas,toth,natario,duztas2,mode,taub-nut,gwak3}. In these experiments the authors obtained similar results for scalar fields however the results for the fermionic fields turned out to be somewhat drastic. The main difference is the occurrence of superradiance, which plays a crucial role in  scattering problems. 

There exists an asymptotically flat black hole solution in the low energy limit of heterotic string theory, which was derived by Sen~\cite{sen}. This rotating charged black hole solution is known as Kerr-Sen black hole. Siahaan applied a test of wCCC on Kerr-Sen black holes with test particles~\cite{siahaan}. He found that test charged particles can drive both extremal and nearly extremal Kerr-Sen black holes beyond the extremal limit, leading to the formation of naked singularities. In this work we check the validity of wCCC in the case of Kerr-Sen black holes interacting with test fields. 
Searching for a counter-example to wCCC in the spirit of Wald type Gedanken experiments, we attempt to overspin Kerr-Sen black holes into naked singularities by neutral test fields. We evaluate the cases of test scalar fields interacting with both nearly extremal and extremal Kerr-Sen black holes. We also ask and answer the question: What if there is no superradiance?
\section{Test field modes on the Kerr-Sen background and wCCC}
In Boyer-Lindquist coordinates, the Kerr-Sen metric can be written as
\begin{equation}
ds^2=-\frac{\Delta}{\Sigma}(dt-a\sin^2 \theta d\phi)^2+\frac{\sin^2 \theta}{\Sigma}[adt-(\Sigma +a^2\sin^2 \theta)d\phi]^2+\Sigma \left(\frac{dr^2}{\Delta}+d\theta^2 \right) \label{kerrsen}
\end{equation}
where $\Delta=r^2+2(b-M)r +a^2 $, and $\Sigma=r^2 + 2br +a^2 \cos^2 \theta$. This metric describes a black hole with mass $M$, charge $Q$, and angular momentum $J=Ma$. The twist parameter $b$ is defined by $b=Q^2/2M$. The inner and outer  horizons are located at
\begin{equation}
r_{\pm}=M-b \pm \sqrt{(M-b)^2-a^2} \label{rplus}
\end{equation}
Notice that the event horizon exists if and only if 
\begin{equation}
(M-b)\geq a  \label{crit}
\end{equation}
where the equality corresponds to the extremal case. The Kerr-Sen metric (\ref{kerrsen}) describes a naked singularity if $M-b-a<0$, since one cannot find a real root for $r_+$. In this work, we check whether a Kerr-Sen black hole initially satisfying (\ref{crit}) can be driven beyond extremality by the interaction with the test fields.

The Klein-Gordon equation for test scalar fields on Kerr-Sen background was studied by Wu and Cai \cite{wu}. The relative scattering probability of the scalar waves at the horizon is found to be \footnote{Equation (15) in \cite{wu} is equivalent to (\ref{scatter}), though a different notation is used by Wu and Cai} 
\begin{equation}
P_{\rm{scattering}}=\exp \left(-2\pi(\omega -m\Omega -q\Phi )/\kappa \right) \label{scatter}
\end{equation}
where $\omega$ is the frequency of the test field, $m$ is the azimuthal wave number, $\Omega=a/2Mr_+$ is the angular velocity of the horizon, $\Phi=Q/2M$ is the electric potential, and $ \kappa=(r_+ -M +b)/2Mr_+ $ is the surface gravity. Though not explicitly stated by Wu and Cai, the scattering probability (\ref{scatter}) becomes larger than 1 if the frequency of the incoming mode is in the range $0<\omega<m\Omega +q\Phi$, i.e. superradiance occurs. Superradiance can roughly be defined as the amplification of waves as they scatter off black holes. Whether or not superradiance occurs is a crucial factor in scattering problems.

\subsection{Scalar fields and nearly extremal Kerr-Sen black holes}
In Wald type Gedanken experiments involving test particles we analyse the geodesic motion. We find the minimum energy for the particle to be absorbed by the black hole. Then we find the maximum energy by demanding that the event horizon is destroyed at the end of the interaction. If there exists a range of energies between the minimum and maximum energy one can conclude that the interaction of the black hole with test particles can lead to a violation of wCCC.

In similar thought experiments involving fields, we envisage a test field incident on the black hole from infinity. The field is partially absorbed by the black hole and partially scattered back. At the end of the interaction the test field decays away, leaving behind a space-time with perturbed parameters. Then, we check if the interaction can drive a black hole beyond extremality.

If we perturb black holes with test fields, we should first demand that the field is absorbed by the black hole  analogous to the particle case. Since superradiance occurs for scalar fields on Kerr-Sen background, the frequency of the incoming field should be larger than the limiting frequency $\omega_{\rm{sl}}$ to ensure that the field is absorbed by the black hole.
\begin{equation}
\omega > \omega_{\rm{sl}}=\frac{ma}{2Mr_+} \label{sl}
\end{equation}
where $r_+$ is given by (\ref{rplus}). Note that (\ref{sl}) is valid for neutral fields. The condition for the event horizon to exist is simply $M-b \geq a$. We can parametrize a nearly extremal black hole as:
\begin{equation}
M-b-a=M\epsilon^2 \Rightarrow 2M^2-Q^2-2J=2M^2\epsilon^2 \label{param}
\end{equation}
Initially the nearly Kerr-Sen space-time satisfies $\delta_{\rm{in}}\equiv 2M^2-Q^2-2J=2M^2\epsilon^2$. After the interaction with the scalar field  the mass and angular momentum parameters of the space-time are perturbed. (The charge parameter is invariant since the scalar field is neutral). We demand that the event horizon is destroyed at the end of the interaction, which gives us  the maximum value for the frequency of the incoming field.
\begin{equation}
\delta_{\rm{fin}}=2(M+\delta E)^2 - Q^2 -2\left( J+ \delta J \right)<0 \label{overspin1}
\end{equation}   
Let us choose $\delta E=M\epsilon$ for the incoming field so that the test field approximation is maintained. Then we substitute $\delta J=(m/\omega)\delta E$  for the scalar field and re-write (\ref{overspin1}), imposing (\ref{param}). Elementary algebra yields that the condition $\delta_{\rm{fin}}<0$ is equivalent to
\begin{equation}
\omega < \omega_{\rm{max}}=\frac{m}{2M(1+\epsilon)} \label{wmax}
\end{equation}
If we perturb a nearly extremal Kerr-Sen black hole with a scalar field with energy $\delta E=M\epsilon$ and frequency in the range $\omega_{\rm{sl}}<\omega<\omega_{\rm{max}}$, the event horizon of the Kerr-Sen black hole is destroyed at the end of the interaction. Apparently this could be possible if $\omega_{\rm{sl}}<\omega_{\rm{max}}$ for the nearly extremal Kerr-Sen black hole parametrized as (\ref{param}). At this stage we have to prove that $\omega_{\rm{sl}}<\omega_{\rm{max}}$, which is not quite apparent in equations (\ref{sl}) and (\ref{wmax}). For that purpose first notice that $r_+$ can be written in the form
\begin{equation}
r_+=a+ M\epsilon^2+\sqrt{(M-b-a)(M-b+a)} \label{rplus1}
\end{equation}
By the parametrization (\ref{param}) $(M-b-a)=M\epsilon^2$. Let us define $(M-b+a)=M\alpha^2$. By definition $\alpha>\epsilon$. Let $\beta$ be the geometric average of $\alpha$ and $\epsilon$, i.e. $\beta^2=\epsilon^2 \alpha^2$. $\beta$ is also larger than $\epsilon$, by definition. We can re-write equation (\ref{rplus1})
\begin{equation}
r_+=a+ M\epsilon^2+ M\beta \label{rplus2}
\end{equation}
The limiting frequency for superradiance can be expressed in the form
\begin{equation}
\omega_{\rm{sl}}=\frac{m}{2M}\left( \frac{a}{r_+} \right)=\frac{m}{2M} \left( \frac{1}{1+(M/a)\epsilon^2 +(M/a)\beta} \right) \label{wsl1}
\end{equation}
Since $\beta>\epsilon$ and $(M/a)>1$, we have $(1+(M/a)\beta) > (1+\epsilon)$, so $(1+ (M/a)\epsilon^2 +(M/a)\beta) > (1+\epsilon)$. Comparing $\omega_{\rm{sl}}$ in the form  (\ref{wsl1}), with $\omega_{\rm{max}}$ given  in (\ref{wmax}), we see that $\omega_{\rm{sl}} < \omega_{\rm{max}}$, independent of the choice of $\epsilon$. Therefore, there exists a range of frequencies $\omega_{\rm{sl}} < \omega< \omega_{\rm{max}}$ for a scalar scalar field to overspin a nearly extremal Kerr-Sen black hole into a naked singularity.

For a numerical example let us consider a nearly extremal Kerr-Sen black hole with $a=0.5M$ and $\epsilon=0.01$ so that $\delta_{\rm{in}}=2M^2-Q^2-2J=0.0002M^2$. We perturb this black hole with a scalar field incident from infinity. We choose the energy of the incident field to be $\delta E=M\epsilon=0.01M$. 
The spatial coordinate of the event horizon can be calculated as $r_+=0.5101005M$. Then the limiting frequency for superradiance is $\omega_{\rm{sl}}=0.4900995(m/M)$, whereas $\omega_{\rm{max}}=0.4950495(m/M)$. We choose the frequency of the incoming field in the range $\omega_{\rm{sl}}<\omega<\omega_{\rm{max}}$. Let us choose $\omega=0.492(m/M)$ for the incoming field. The contribution of this field to the angular momentum parameter of black hole is $\delta J= (m/\omega)\delta E=0.0203252M^2$. We can calculate $\delta_{\rm{fin}}$
\begin{equation}
\delta_{\rm{fin}}=2(M+ \delta E)^2 - Q^2 -2(J+\delta J)=-0.0002504M^2 \label{deltafin}
\end{equation}
The negative sign for $\delta_{\rm{fin}}$ indicates that the event horizon of the Kerr-Sen black hole is destroyed after the interaction with the scalar field. Thus, wCCC is violated ignoring backreaction effects. Previously, we derived that there exists a narrow range frequencies above the superradiance limit (\cite{overspin}), that can be used to overspin a nearly extremal Kerr black hole. The analysis for Kerr-Sen black holes yields an analogous result.
\subsection{Scalar fields and extremal black holes}
An extremal Kerr-Sen black hole satisfies 
\begin{equation}
M-b-a=0\Rightarrow 2M^2-Q^2-2J=0 \label{paramex}
\end{equation}
In the case of nearly extremal black holes we both parametrized $\delta_{\rm{in}}$ and the energy of the incoming field by the same number $\epsilon$. For extremal black holes $\delta_{\rm{in}}=0$ by definition. We choose the energy of the incoming field $\delta E=M\epsilon^{\prime}$, where $\epsilon^{\prime}<<1$. The limiting frequency for superradiance is given by
\begin{equation}
\omega_{\rm{sl-ex}}=\frac{m}{2M} \label{wslex}
\end{equation}
which directly follows from the fact that $r_+=a$ for extremal Kerr-Sen black holes. Again we demand that the extremal black hole is overspun at the end of the interaction, to find the max value for  the frequency of the incoming field. 
\begin{equation}
\delta_{\rm{fin}}=2(M+M\epsilon^{\prime})^2 -Q^2 -2(J+(m/\omega)M\epsilon^{\prime})<0
\end{equation}
where we have made the substitutions $\delta E=M\epsilon^{\prime}$ and $\delta J=(m/\omega)\delta E$. Then the condition $\delta_{\rm{fin}}<0$ is equivalent to
\begin{equation}
\omega<\omega_{\rm{max-ex}}=\frac{m}{M(\epsilon^{\prime}+2)}
\end{equation}
Apparently $\omega_{\rm{max-ex}}$ is less than the superradiant limit for extremal Kerr-Sen black holes given in (\ref{wslex}). The test scalar fields that could possibly overspin an extremal Kerr-Sen black hole are not absorbed by the black hole due to superradiance. Therefore scalar fields cannot overspin extremal Kerr-Sen black holes analogous to the Kerr case. This analogy is not trivial. In particular, we have recently shown that extremal Kerr-Taub-Nut black holes can be overspun by scalar fields \cite{taub-nut}.
\subsection{Fermionic fields and the absence of superradiance}
Whether or not the fermionic fields on Kerr-Sen background exhibit superradiant  scattering is an open problem. Yet, from Kerr analogy one can naively assume that superradiance does not occur for fermionic fields. In such a case there would not be a lower limit for the frequency of the incoming field. The absorption of modes carrying low energy and relatively high angular momentum would be possible. Then extremal Kerr-Sen black holes could also be overspun since the absorption of the modes with $\omega<\omega_{\rm{sl-ex}}$ would be allowed.  Overspinning of both extremal and nearly extremal black holes would become generic since the absorption of modes with low energy and high angular momentum leads to a considerable increase in the absolute value of $\delta_{\rm{fin}}$ at the end of the interaction.

To clarify our arguments let us reconsider the example for nearly extremal black holes, where $\delta_{\rm{in}}=0.0002M^2$. This time we perturb the black hole by a fermionic field with frequency $\omega=0.25 (m/M)$. This field will be absorbed by the black hole if there is no superradiance. The contribution of this field to the angular momentum parameter is much higher $\delta J=(m/\omega)\delta E=0.04M^2$. We can calculate $\delta_{\rm{fin}}$
\begin{equation}
\delta_{\rm{fin}}=2(M+ \delta E)^2 - Q^2 -2(J+\delta J)=-0.0396M^2 \label{deltafinferm}
\end{equation}
We see that $\vert\delta_{\rm{fin}}\vert$ is not of the order of $~M^2\epsilon^2$ as in the case of scalar fields. The overspinning of the Kerr-Sen black holes by fermionic fields is more robust compared to scalar fields. We should note that the arguments in this section are based on the assumption that superradiance does not occur for fermionic fields on Kerr-Sen background. 

Actually, the absence of superradiance leads to drastic results challenging the validity of cosmic censorship. Previous analysis for Kerr \cite{duztas2,superrad}, BTZ \cite{btz}, and Kerr-Taub-NUT \cite{taub-nut} backgrounds, also imply a generic violation of wCCC if there is no superradiance.
\section{Conclusions}
In this work we attempted to overspin Kerr-Sen black holes with neutral test fields. We envisaged test fields incident on the black hole from infinity with frequency $\omega$ and azimuthal wave number $m$. Though it was not explicitly stated, the  previous study of scalar fields on Kerr-Sen background by Wu and Cai implies that superradiance occurs for scalar fields analogous to the Kerr case. Therefore the frequency of the incoming field should be larger than the superradiance limit to ensure that the field is absorbed by the black hole. Otherwise, the wave is scattered back with a larger amplitude borrowing the access energy from the rotational energy of the black hole. The angular momentum parameter of the black hole decreases. Superradiant scattering of test fields drives the black hole away from extremality, which reinforces wCCC. For that reason the minimum frequency to ensure the absorption of a scalar field is equivalent to the superradiance limit.  We derived the maximum frequency by demanding that the black hole is overspun into a naked singularity at the end of the interaction. We showed that the maximum frequency is larger than the superradiance limit for nearly extremal black holes. Therefore, there exists a range of frequencies that can be used to drive a nearly extremal black hole beyond extremality. However, no such range exists for extremal Kerr-Sen black holes. Extremal Kerr-Sen black holes cannot be overspun by test scalar fields.

We also evaluated the case of no superradiance, which is a  naive expectation from Kerr analogy. In this case, both extremal and nearly extremal black holes can be generically overspun. For the case of scalar fields we found that $ \delta_{\rm{fin}}\sim -M^2 \epsilon^2 $, which indicates that the destruction of the horizon can possibly be fixed by the backreaction effects ignored in this work. However, if superradiance does not occur, the modes carrying low energy and high angular momentum can be absorbed. In that case $\vert \delta_{\rm{fin}}\vert \gg M^2 \epsilon^2 $, which represents a robust violation of wCCC that cannot be fixed by employing backreaction effects. This is analogous to the Kerr case.

\end{document}